\documentstyle[twoside,fleqn,espcrc2]{article}
\title{Generalized canonical quantization of bosonic string
model in massive background fields}
\author{I.L. Buchbinder\address{Department of Theoretical Physics,
Tomsk State Pedagogical University,\\ Tomsk 634041, Russia},
V.D. Pershin\address{Department of Theoretical Physics,
Tomsk State University,\\ Tomsk 634050, Russia}
and G.B. Toder\address{Department of Physics and Chemistry,
 Omsk State Academy of Railway Communications,\\
 Omsk 644046, Russia}}
\begin{document}

\begin{abstract}
A method of constructing a canonical gauge invariant quantum
formulation for a non-gauge classical theory depending on a
set of parameters is advanced
and then applied to the theory of closed bosonic string
interacting with  massive background fields.
It is shown that within the proposed formulation
the correct linear equations of motion for background fields arise.
\end{abstract}

\maketitle

\section{Introduction}
BFV method is the most powerful realization of canonical
quantization procedure which provides unitarity at the
quantum level and consistency of a theory symmetries and dynamics.
Now it has been studied in details \cite{BFV,Hen} but
formulation of new models in quantum field theory requires
investigation of its yet unexplored aspects..

One of these aspects arise from bosonic string theory coupled to
background fields \cite{callan}. At the quantum level string models
should be conformal invariant and this requirement leads to
restrictions on spacetime dimension in the case of free string
theory and to effective equations of motion
for massless background fields in the case of string theories
coupled to background \cite{callan} (see also the reviews
\cite{tseyt}).  In terms of covariant functional methods this
condition appears as independence of quantum effective action
on the conformal factor of two-dimensional metrics or as
vanishing of renormalized operator of the energy-momentum tensor
trace.

According to the prescription \cite{Pol} generally accepted in
functional approaches to string theory dynamical
variables should be treated differently. Namely,
functional integration is carried out only over string
coordinates $X^\mu(\tau,\sigma)$ while components of
two-dimensional metrics $g_{ab}(\tau,\sigma)$ are considered as
external fields. Then one demands the result of such an
integration to be independent on the conformal factor and the
integrand over $g_{ab}(\tau,\sigma)$ reduces to finite
dimensional integration over parameters specifying string world
sheet topologies. This prescription differs from the standard
field theory rules when functional integral is calculated over
every variable independently. In string theory such an
independent integration would lead to appearence at the quantum
level of an extra degree of freedom connected with
two-dimensional gravity \cite{buchshap}.

This approach can also be applied to string theory
interacting with massive background fields which is not
classically conformal invariant \cite{ovrut}. One demands that
operator of the energy-momentum tensor trace vanish no matter whether
the corresponding classical action is conformal invariant or not. As
was shown in \cite{buchper,krykh} it gives rise to effective
equations of motion for massive background fields.
Unfortunately, in the case of closed string theory
covariant approaches did not reproduce the full set of correct
linear equations of motion for massive background fields.
Namely, the tracelessness condition on massive
tensor fields has not been obtained neither in
standard covariant perturbation approaches \cite{buchper} nor
within the formalism of exact renormalization group \cite{ellw}.
So there exists a problem how to derive the correct
equations for massive background fields.

Moreover, from general point of view the requirement of quantum
Weyl invariance of string theory with massive background fields
means that a non-gauge classical theory depending on a set of
parameters is used for constructing of a quantum theory that is
gauge invariant under some special values of the parameters.
Such a situation occurs in string theory if interaction with
massless dilaton, tachyon or any other massive field is turned
on. As we consider canonical approach to be the only completely
consistent method for constructing quantum theories then every step
of any quantization procedure should be justified by an
appropriate prescriptions within canonical formulation.  So the
general problem arising from string theories is how to describe
in terms of canonical quantization construction of gauge
invariant quantum theory starting with a classical theory
without this invariance.

Due to the general BFV method one
should construct hamiltonian formulation of classical theory,
find out all constraints and calculate algebra of their Poisson
brackets.  Then one defines fermionic functional $\Omega$
generating algebra of gauge transformations and bosonic
functional $H$ containing information of theory dynamics.
Quantum theory is consistent provided that the operator $\hat\Omega$
is nilpotent and conserved in time. The corresponding analysis
for bosonic string coupled to massless background fields was
carried out in \cite{buch91,buch95}.

In the case of string theory interacting with massive background
fields components of two-dimensional metrics should be treated
as external fields, otherwise classical equations of
motion would be inconsistent. As a consequence, classical
gauge symmetries are absent and it is impossible to construct
classical gauge functional $\Omega$. In this paper we consider a
prescription \cite{toder} allowing for some models to construct
quantum operator $\hat\Omega$ starting with a classical theory
without first class constraints. Our method do not assume
introduction of extra degrees of freedom for restoring gauge
invariance at the classical level. Classical theory within our
approach remains non-gauge but quantum theory is gauge invariant if
there exist values of theory parameters providing nilpotency and
conservation of operator $\hat\Omega$. To illustrate how
the prescription works we apply it to the theory of closed
bosonic string coupled to masssive background fields and show
that correct linear equations of motion are produced.

\section{General prescription}
Consider a system described by a Hamiltonian
\begin{equation}
H=H_{0}(a)+\lambda^{\alpha}T_{\alpha}(a)
\label{Hgen}
\end{equation}
where $H_0(a)=H_0(q,p,a)$, $T_{\alpha}(a)=T_{\alpha}(q,p,a)$ and
$q$, $p$ are canonically conjugated dynamical variables;
$a=a_i$ and $\lambda^{\alpha}$ are external parameters of the
theory.

We suppose that $T_{\alpha}(a)$ are some functions of the form
\begin{equation}
T_{\alpha}(a)=T^{(0)}_{\alpha}(a)+T^{(1)}_{\alpha}(a)
\end{equation}
and closed algebra in terms of Poisson brackets is formed by
$T^{(0)}_{\alpha}(a)$, not by $T_{\alpha}(a)$:
\[
\{ T^{(0)}_{\alpha}(a),T^{(0)}_{\beta}(a) \}  =
T^{(0)}_{\gamma}(a)U^{\gamma} _{\alpha\beta}(a)
\]
\begin{equation}
\{ H_0(a),T^{(0)}_{\alpha}(a) \} =
T^{(0)}_{\gamma}(a)V^{\gamma} _{\alpha}(a)
\end{equation}
Such a situation may occur, for example, if
$T^{(0)}_{\alpha}(a)$ correspond to a free gauge invariant
theory and $T^{(1)}_{\alpha}(a)$ describe a perturbation
spoiling gauge invariance. At the quantum level both the algebras of
$T^{(0)}_{\alpha}(a)$ and $T_{\alpha}(a)$ are not closed in
general case.

We define quantum operators $\Omega$ and  $H$ as follows:
\[
\Omega  =
c^{\alpha}T_{\alpha}(a)-{1\over2}U^{\gamma}_{\alpha\beta}(a)
:{\cal P_{\gamma}} c^{\alpha} c^{\beta}:
\]
\begin{equation}
H  =  H_0 (a) + V^{\gamma}_{\alpha}(a) :{\cal P_{\gamma}}
c^{\alpha}:
\label{Omega}
\end{equation}
where $:\quad:$ stands for some ordering of ghost fields.

In general case $\Omega^2 \neq 0$ and $d\Omega/dt \neq 0$.
However, if there exist some specific values of parameters $a$ that make
the operator $\Omega$ to be nilpotent and conserved  then the corresponding
quantum theory is gauge invariant. Thus it is possible
to construct a quantum theory with given gauge
invariance that is absent at the classical level.

\section{String theory in massive fields}
As  an example where the described procedure really
works and leads to nilpotency and conservation conditions for
the operator $\Omega$  with non-trivial solutions for parameters $a$
we consider closed bosonic string theory coupled with background
fields of tachyon and of the first massive level. We will restrict
ourselves by linear approximation in background fields because an
adequate treatment of non-linear (interaction) terms is known to
demand non-perturbative methods \cite{Das}.  This approximation will
be enough to establish consistency of background fields dynamics with
structure of the corresponding massive levels in string spectrum.

The theory is described by the classical action
\[
S={}-{1\over 2\pi\alpha'} \int d^{2}\sigma\,
  \sqrt{-g} \Bigl\{{1\over 2} g^{ab}\partial_{a} X^{\mu} \partial_{b}
  X^{\nu} \eta_{\mu\nu}
\]\[
{}+  g^{ab} g^{cd} \partial_a X^\mu \partial_b X^\nu
          \partial_c X^\lambda \partial_d X^\kappa
        F^1_{\mu\nu\lambda\kappa}(X)
\]\[
{}+ g^{ab}\varepsilon^{cd} \partial_a X^\mu \partial_b X^\nu
          \partial_c X^\lambda \partial_d X^\kappa
      F^2_{\mu\nu\lambda\kappa}(X)
\]\[
{}+ \alpha' R^{(2)} g^{ab} \partial_a X^\mu \partial_b X^\nu
      W^1_{\mu\nu}(X)
\]\[
{}+ \alpha' R^{(2)} \varepsilon^{ab} \partial_a X^\mu \partial_b
X^\nu W^2_{\mu\nu}(X)
\]
\begin{equation}
{}+  \alpha'^2 R^{(2)} R^{(2)} C(X) + Q(X)\Bigr\},
\label{S}
\end{equation}
$\sigma^a=(\tau,\sigma)$ are coordinates on string world sheet,
$R^{(2)}$ is scalar curvature of two-dimensional  metrics
$g^{ab}$, $\eta_{\mu\nu}$ is Minkowski metrics of
$D-$dimensional spacetime,  $Q$ is tachyonic field and $F$,
$W$, $C$ are background fields of the first massive level in
string spectrum.  As was shown in \cite{buchper} all other
possible terms with four two-dimensional derivatives in
classical action are not essential and string interacts with
background fields of the first massive level only by means of
the terms presented in (\ref{S}).

Components of two-dimensional metrics $g_{ab}$ should be
considered as external fields, otherwise the classical equations
of motion $\delta S/\delta g_{ab}=0$ would be fulfilled only for
vanishing background fields. This treatment is similar
to covariant methods where functional integral
is calculated only over $X^\mu$ variables.

After the standard parametrization of metrics
\begin{eqnarray}
g_{ab}& = &e^{\gamma}\left(\begin{array}{cc}
\lambda^{2}_{1}-\lambda^{2}_{0} & \lambda_{1} \\
\lambda_{1}  & 1 \end{array}\right)
\end{eqnarray}
the Hamiltonian in linear approximation in background fields
takes the form
\begin{equation}
H=\int d\sigma \,(\lambda_{0}T_{0}+\lambda_{1}T_{1}),
\label{H}
\end{equation}
where
\[
T_0 = T_0^{(0)} + T_0^{(1)}, \quad T_1 = T_1^{(0)} = P_{\mu}X'^\mu {,}
\]\[
T_0^{(0)} = (1/2) \Bigl(2\pi\alpha'P^2 + (2\pi\alpha')^{-1} X'^2
\Bigr),
\]\[
T_0^{(1)} = (1/2\pi) \Bigl( (\alpha')^{-1}e^{\gamma} Q
\]\[
{}+ (\alpha')^{-1} e^{-\gamma}
Y^{+\mu}Y^{+\nu}Y^{-\lambda}Y^{-\kappa} F_{\mu\nu,\lambda\kappa}
\]
\begin{equation}
{} + R^{(2)} Y^{+\mu}Y^{-\nu} W_{\mu\nu}
+ \alpha' e^\gamma R^{(2)} R^{(2)} C \Bigr) ,
\label{T}
\end{equation}
$P_\mu$ are momenta canonically conjugated to $X^\mu$,
$X'^\mu=\partial X^\mu/\partial\sigma$
and we introduced the following notations:
\[
Y^{\pm\mu} = 2\pi\alpha' P^\mu \mp X'^\mu, \quad
W_{\mu\nu}=-W^1_{\mu\nu}+W^2_{\mu\nu},
\]
\begin{equation}
F_{\mu\nu,\lambda\kappa} =
2 F^1_{\mu\lambda\nu\kappa} + 2 F^1_{\mu\kappa\nu\lambda} -
2 F^2_{\mu\lambda\nu\kappa} - 2 F^2_{\nu\kappa\mu\lambda}
\end{equation}
$T_0^{(0)}$ and $T_1^{(0)}$ represent constraints of free string
theory and form closed algebra in terms of Poisson brackets.
$\lambda_0$ and $\lambda_1$ play the role of external fields and
so $T_0$ and  $T_1$ cannot be considered as constraints of
classical theory. In free string theory conditions
$T^{(0)}_0=0$, $T^{(0)}_1=0$ result from conservation of
canonical momenta conjugated to $\lambda_0$ and $\lambda_1$.
According to our prescription in string theory with massive
background fields  $\lambda_0$ and $\lambda_1$ can not be
considered as dynamical variables, there are no corresponding
momenta and conditions of their conservation do not appear.

The role of parameters $a$ in the theory under consideration is
played by background fields $Q$, $F$, $W$, $C$ and conformal
factor of two-dimensional metrics $\gamma(\tau,\sigma)$. The
theory (\ref{H}) is of the type (\ref{Hgen}) with $H_0=0$,
structural constants of classical algebra being independent on
time.

Direct calculations up to terms linear in background fields show
that the operator $\Omega$ defined according to (\ref{Omega})
is nilpotent and conserves provided that the following conditions are
fulfilled:
\[ D=26, \quad \gamma=const, \]\[ (\partial^2 +
4/\alpha')Q=0,\quad (\partial^2 -
4/\alpha')F_{\mu\nu,\lambda\kappa}=0 , \]\[ \partial^\mu
F_{\mu\nu,\lambda\kappa}=0, \quad \partial^\lambda
F_{\mu\nu,\lambda\kappa}=0, \]
\begin{equation}
F^\mu{}_{\mu,\lambda\kappa}=0, \quad
F_{\mu\nu,}{}^\lambda{}_\lambda=0.
\label{shell}
\end{equation}
The condition $\gamma=const$ means that
string world sheet should be flat $R^{(2)}=0$ and so
the background fields $W$ and $C$ disappear from the classical
action (\ref{S}). The eqs.(\ref{shell}) show
that first massive level is described by a tensor
of fourth rank which is symmetric and traceless in two pairs of
indices and transverse in all indices. This exactly corresponds
to the closed string spectrum and so our approach gives the full
set of correct linear equations for massive background fields.

The described example demonstrates a possibility to construct
canonical formulation of quantum theory invariant under gauge
transformations that are absent at the classical level. The
proposed method opens up a possibility for deriving interacting
effective equations of motion for massive and massless
background fields within the framework of canonical formulation
of string models and provides a justification of covariant
functional approach to string theory.

\section*{Acknowledgments}
The authors are grateful to E.S.~Fradkin, M.~Henneaux, P.M..~Lavrov,
R.~Marnelius, B.~Ovrut, A.A.~Tseytlin and I.V.~Tyutin for
discussions of some aspects of the paper. The work was supported
by RFBR, project No.~96-02-16017 and RFBR-DFG, project
No.~96-02-00180G.

\end{document}